# Homogeneous and Non Homogeneous Algorithms

**Paparrizos K. Ioannis**

Department of Informatics
Aristotle University of Thessaloniki
54124 Thessaloniki, Greece
E-mail: ipapa@csd.auth.gr

**Abstract**
Motivated by recent best case analyses for some sorting algorithms and based on the type of complexity we partition the algorithms into two classes: homogeneous and non homogeneous algorithms. Although both classes contain algorithms with worst and best cases, homogeneous algorithms behave uniformly on all instances. This partition clarifies in a completely mathematical way the previously mentioned terms and reveals that in classifying an algorithm as homogeneous or not best case analysis is equally important with worst case analysis.
**Keywords:** Algorithm analysis, Algorithm complexity, Algorithm classification.

## 1. Introduction

In the 70s and 80s a lot of discussion was going on regarding the right use of the asymptotic symbols Ο, Θ and Ω used to analyze algorithms and compare their theoretical efficiency. Some researchers use these symbols to denote the rate of growth of functions and others to denote sets of functions; see relevant comments in [Brassard (1985)], [Gurevich (1986)] and [Knuth (1976)]. Following the approach of using the asymptotic symbols as sets of functions we partition the class of algorithms into two non empty subclasses: *homogeneous* and *non homogeneous* algorithms. Both classes are wide. They contain iterative and recursive algorithms. Although both classes contain algorithms with worst and best cases, homogeneous algorithms behave uniformly on all instances of the problem being solved. The partition clarifies in a completely mathematical way the terms of algorithm, worst and best case complexity, the only difference between them being the sets of instances they referred to.

This classification of algorithms was triggered by recent theoretical result concerning best case analysis of some heapsort algorithms [Bollobas et.al (1996)], [Ding et.al (1992)], [Dutton (1993)], [Edelkamp (2000)], [Edelkamp (2002)], [Fleischer (1994)] [Schaffer et.al (1993)] and [Wang et.al (2007)]. Also, computational results indicate that best case analysis might have practical value too, see for example [Edelkamp (2002)] and [Wang et.al (2007)]. Our results indicate that in order to classify an algorithm as homogeneous or not the complexity of the exact, up to a set of functions defined by the asymptotic symbol Θ, best case and worst case must be computed.

When the classification is accomplished the analysis of the complexity of the algorithm is complete, indicating, from a theoretical point of view, that best case analysis is equally important with the worst case analysis.

The term inhomogeneity has been used by Nadel B.A. [Nadel (1993)] who characterises the imprecision of an analysis of an algorithm in terms of the difference $\Delta_C = c_w - c_b$ between the worst and best case complexity, where C is a proper measure of complexity. In particular, for the sorting problem C is the number of comparisons. Using various combinations of disorder parameters, Nadel [Nadel (1993)] partitions the set of instances in big, medium, small, tiny and singleton subclasses and computes the inhomogeneity in each subclass. Other relevant results for other problems are presented in [Haralick et.al (1980)],[Nadel (1986)],[Nadel (1990)] and [Nudel (1983)]. Our approach is different in the sense that the set of algorithms is partitioned and not the set of instances of the problem.

In the next section we formally describe the two classes of algorithms. Some details regarding the algorithm classification are presented in section 3. Recursive and divide and conquer homogeneous and non homogeneous algorithms are discussed and some side results are also presented in the last section.

## 2. Description of the two classes.

We derive our results using the *Random Access Machine (RAM)* model in which every elementary operation such as addition, subtraction, multiplication and division of two numbers, comparison of two numbers, reading and writing a number in the memory, calling a function, etc., is executed in constant time. It is well known that all constant functions belong to the set $\Theta(1)$. Recall that $\Theta(g(n))$ denotes a set of functions defined as follows:

**Definition 1**. Given a function $g(n)$ we denote by $\Theta(g(n))$ the set of functions $t(n)$ for which there exists constants $a > 0$ and $b > 0$ and a positive integer $n_0$ such that

$$b\, g(n) \leq t(n) \leq a\, g(n) \tag{1}$$

for every $n \geq n_0$.

All functions used in this paper denote time and therefore they are positive. The argument n denotes the dimension of the problem and, hence, it is a positive integer.

The sets of functions $O(g(n))$ and $\Omega(g(n))$ are similarly defined. Simply, in the definition of $O(g(n))$ the left inequality of (1) is missing while in the definition of $\Omega(g(n))$ the right. Observe that $\Theta(g(n))$ is strictly contained in the sets $O(g(n))$ and $\Omega(g(n))$. As a result the assumption that the basic operations are executed in $\Theta(1)$ time (instead of $O(1)$ or $\Omega(1)$ time) provides a more precise algorithm analysis.

It is well known that the symbol $\Theta$ considered as a binary relation between functions is reflexive, symmetric and transitive and therefore it partitions the set of functions into disjoined classes. In other words, if f(n) and g(n) are two different functions, then either $\Theta(f(n)) = \Theta(g(n))$ or $\Theta(f(n)) \cap \Theta(g(n)) = \emptyset$. In particular the following two results are well known.

**Theorem 1.** If $f(n) \in \Theta(g(n))$, then $\Theta(f(n)) = \Theta(g(n))$.

**Theorem 2.** The sets $\Theta(1)$ and $\Theta(n)$ are disjoint.

Given a computational problem we denote the set of instances of dimension n by I(n). Consider now an algorithm A solving the problem under consideration. The time taken by algorithm A to solve instance i of dimension n is denoted by $t_A(i, n)$. In algorithm analysis we try to describe in a nice way the set of time functions

$$S = \{ t_A(i, n) : i \in I(n) \}$$

One way to do this is via the sets of functions defined by the asymptotic symbols O, $\Theta$, $\Omega$. We are completely satisfied if we can determine a function g(n) such that

$$S \subseteq \Theta(g(n)) \quad (2)$$

Once again, observe that we use the set $\Theta(g(n))$ which is strictly contained in the sets $O(g(n))$ and $\Omega(g(n))$ and therefore the description of set S is more precise. This preference though leads us naturally to the following definition.

**Definition 2**. An algorithm is *homogeneous* if there exists a function g(n) such that relation (2) holds. Otherwise, the algorithm is *non homogeneous*.

**Theorem 3.** The class of algorithms is partitioned into two non empty and disjoined subclasses, the subclasses of homogeneous and non homogeneous algorithms.

**Proof.** Let U be the class of all algorithms, H the class of homogeneous and NH the class of non homogeneous algorithms. It is obvious from Definition 2 that

$$H \cap NH = \emptyset \quad \text{and} \quad H \cup NH = U.$$

It remains to show that $H \neq \emptyset$ and $NH \neq \emptyset$. This proof is done by providing a simple algorithm for each class.

Firstly, consider the problem of finding the smallest among n given numbers stored as elements of an array T. The algorithm *min* whose pseudocode is the following

```
a←T(1)
for j = 2, 3,…, n
    if T(j) < a,  a←T(j),
```

solves this problem and is homogeneous. Indeed assuming that an element of an array is reached in constant time $\Theta(1)$ in the computational model of constant times, it is easy to conclude that

$$t_{min}(i, n) \in \Theta(n)$$

for every instance $i \in I(n)$. Hence, algorithm min is homogeneous and $H \neq \emptyset$.

Secondly, consider the following problem. Given an array T of n elements sorted in increasing order, i.e.

$$T(j) \leq T(j+1) \quad \text{for} \quad i = 1, 2, \ldots, n-1,$$

and a number x, sort all elements of T and the number x in increasing order.

This problem is solved by the algorithm *insert* whose pseudocode is the following:

$j \leftarrow n$, $T(n+1) \leftarrow x$

**while** $(j \geq 1)$ **and** $(T(j) > T(j+1))$

$\quad$ temp$\leftarrow T(j)$, $T(j) \leftarrow T(j+1)$, $T(j+1) \leftarrow$ temp

$\quad j \leftarrow j-1$

Denote by $i_b$ the instance $T = [1\ 2\ 3\ \ldots\ n\text{-}1\ n]$ and $x = n+1$. When algorithm *insert* is applied on instance $i_b$, the while loop is executed once and hence,

$$t_{insert}(i_b, n) \in \Theta(1). \tag{3}$$

Denote now by $i_w$ the instance $T = [1\ 2\ 3\ \ldots\ n\text{-}1\ n]$ and $x = 0$. When algorithm *insert* is applied on instance $i_w$, the while loop is executed $\Theta(n)$ times and therefore

$$t_{insert}(i_w, n) \in \Theta(n). \tag{4}$$

This is so because the first two assignments of the pseudo code insert are executed in $\Theta(1)$ time and each execution of the while loop takes $\Theta(1)$ time.

We show now that there is no function g(n) such that relation (2) holds. This in turn shows that algorithm insert is non homogeneous. Suppose on the contrary that such a function g(n) does exist. By relation (2) we conclude that

$$t_{insert}(i_b, n) \in \Theta(g(n)) \quad \text{and} \quad t_{insert}(i_w, n) \in \Theta(g(n)) \tag{5}$$

By Theorem 1 and relations (3) and (4) we conclude that

$$\Theta(t_{insert}(i_b, n)) = \Theta(1) \quad \text{and} \quad \Theta(t_{insert}(i_w, n)) = \Theta(n) \tag{6}$$

Combining Theorem 1 and relations (5) we conclude that

$$\Theta(t_{insert}(i_b, n)) = \Theta(t_{insert}(i_w, n)) = \Theta(g(n)) \tag{7}$$

Finally, from relations (6) and (7) we conclude that $\Theta(1) = \Theta(n)$, which contradicts Theorem 2. This completes the proof of the Theorem.

In the proof of Theorem 3 we used two simple algorithms to show that the classes of homogeneous and non homogeneous algorithms are non empty. In fact both classes are wide and include recursive and iterative algorithms. The class of non homogeneous algorithms includes plenty of iterative algorithms. The great majority of recursive and divide and conquer algorithms are homogeneous. Among the exceptions is the well known recursive sorting algorithm quick sort [Hoare (1962)] and Euclid's algorithm for computing the greatest common divisor of two numbers.

## 3. Algorithm classification

The instances $i_b$ and $i_w$ used in Theorem 3 are the well known best and worst cases respectively. We call $i_b$ *minimum time instance* and $i_w$ *maximum time instance*. More precisely, we give the following definition.

**Definition 3**. An instance i is a *minimum (maximum) time instance* for an algorithm A, if the total number of elementary operations executed when algorithm A is applied on it is the minimum (maximum) possible.

The analysis so far and particularly algorithm min used in the proof of Theorem 3 might mislead someone to conclude that homogeneous algorithms do not contain minimum and maximum time instances. This is not correct. A striking example of an iterative homogeneous algorithm containing minimum and maximum time instances is the well known Floyd's classical algorithm [Floyd (1964)] for building an initial heap. A heap is a data structure introduced in [Williams (1964)] to develop an efficient general iterative sorting algorithm known today as heapsort. A recursive homogeneous algorithm containing worst and best cases is the well known algorithm in [Blum et.al (1973)], which computes order statistics in linear time.

Some algorithms are obviously homogeneous. If this is not clear for a new algorithm with unknown complexity, using Definition 3 we can set

$$S_b = \{ i_b: i_b \in I(n) \text{ is a minimum time instance}\},$$

$$S_w = \{ i_w: i_w \in I(n) \text{ is a maximum time instance}\}.$$

In the worst (best) case analysis of an algorithm we try to determine a set $\Theta(g(n))$ ($\Theta(f(n))$) containing the set $S_w$ ($S_b$) and say that the worst (best) case complexity of the algorithm is $\Theta(g(n))$ ($\Theta f(n))$). Observe the similarities among the worst and best case complexities of a non homogeneous algorithm and the complexity of a homogeneous algorithm. In particular, the only difference is the set of instances on which they are referred to. Therefore, all these complexities should be described by sets of the form $\Theta(g(n))$.

It is now of interest to determine the complexity of a non homogeneous algorithm, i.e, to find a set of functions including set S. Since a set of the form $\Theta(g(n))$ does not exist, we generalize Definition 1 as follows.

**Definition 4**. Given two (proper) functions $f(n)$ and $g(n)$ we denote by $\Theta(f(n), g(n))$ the set of functions $t(n)$ for which there exist constants $a > 0$ and $b > 0$ and a positive integer $n_0$ such that

$$bf(n) \le t(n) \le ag(n)$$

for $n \ge n_0$.

It is easy to see that $\Theta(f(n), g(n)) = \Omega(f(n)) \cap O(g(n))$. It is also easy to see that the sets $\Theta(0, \infty) = \Omega(0) = O(\infty)$ include always set S. However, in order to be as precise as possible, we are always looking for a minimal set containing set S. In the case of non homogeneous algorithms we are seeking the minimal set $\Theta(f(n), g(n))$ containing set S. Obviously, the set $\Theta(f(n), g(n))$ is minimal if there exist worst and best case instances $i_w$ and $i_b$ such that $t(i_w, n) \in \Theta(g(n))$ and $t(i_b, n) \in \Theta(f(n))$, respectively. Recall that the set $\Theta(1, n)$ describing the complexity of algorithm insert in Theorem 3 is minimal. Observe also that the classification of an algorithm as homogeneous or not is not possible unless the set $\Theta(f(n), g(n))$ describing its complexity is minimal. As the set $\Theta(f(n), g(n))$ is described by best and worst case complexities, both complexities are equally important from the theoretical point of view.

## 4. *Additional results and discussion*

We mentioned earlier that homogeneous algorithms contain worst and best cases. Hence, the average complexity of a homogeneous algorithm is easily defined. Clearly, the mean time of the algorithm on a random instance is

$$t(n) = \frac{\sum_{i \in I(n)} t(i, n)}{|I(n)|},$$

where $|I(n)|$ denotes the number of elements of set $I(n)$. If the complexity of the homogeneous algorithm is $\Theta(g(n))$, it is natural to expect that $t(n) \in \Theta(g(n))$. Indeed, this is the case.

**Theorem 4.** The average complexity of a homogeneous algorithm of complexity $\Theta(g(n))$, is also $\Theta(g(n))$.

**Proof.** Let $t(n)$ be the expected time to solve a random instance. Then

$$t(n) = \frac{\sum_{i \in I(n)} t(i, n)}{|I(n)|} \in \frac{\sum_{i \in I(n)} \Theta(g(n))}{|I(n)|} = \frac{|I(n)| \Theta(g(n))}{|I(n)|} = \Theta(g(n))$$

and the proof is complete.

Observe that this result is independent of the distribution of the instances.

So far we focused our attention on iterative algorithms. Recursive algorithm can be homogeneous and non homogeneous too. But how recursive homogeneous and non homogeneous algorithms look like? A recursive or divide and conquer algorithm makes a fixed number of calls to itself. Therefore, if each call is made on a problem with fixed dimension, the algorithm is homogeneous provided the work required to solve all sub-problems dominates the remaining work. On the contrary, if the dimensions of the sub problems on which calls are made are not fixed and depend on the instance, the algorithm might very well be non homogeneous. Recall that this is the case for the algorithm quicksort [Hoare (1962)]. A recursive or divide and conquer algorithm can be non homogeneous if the number of calls to sub problems is not fixed and depends on the instance. This is the case for Euclid's algorithm computing the greatest common divisor.

## 5. Acknowledge

We thank an anonymous referee for useful suggestions and for bringing to our attention the reference [Nadel (1993)]